\pdfoutput=1
\documentclass[letterpaper, twocolumn]{revtex4}

\usepackage{natbib}

\usepackage{url}
\usepackage{amsmath}
\usepackage{graphicx}
\usepackage{subfigure}

\newcommand{\br}{{\mathbf{r}}}
\newcommand{\bF}{{\mathbf{F}}}
\newcommand{\btheta}{{\mbox{\boldmath$\theta$}}}
\newcommand{\bthetaSmall}{{\mbox{\boldmath$\scriptstyle\theta$}}}

\usepackage{chapterbib}

\begin{document}










\title{Sloppiness, robustness and evolvability in systems biology}


\author{Bryan C. Daniels}
\author{Yan-Jiun Chen}
\author{James P. Sethna}

\address{Laboratory of Atomic and Solid State Physics, Cornell University,
Ithaca, NY, USA}
\author{Ryan N. Gutenkunst}
\address{Department of Biological Statistics and Computational Biology, 
Cornell University, Ithaca, NY, USA}
\author{Christopher R. Myers}
\address{Computational Biology Service Unit, Life Sciences Core Laboratories
Center, Cornell University, Ithaca, NY, USA}	

\begin{abstract}
The functioning of many biochemical networks is often robust --
remarkably stable under changes in external conditions and internal
reaction parameters. Much recent work on robustness and
evolvability has focused on the structure of neutral spaces, in which
system behavior remains invariant to mutations. 
Recently we have shown that the
collective behavior of multiparameter models is most often {\em
sloppy}: insensitive to changes except along a few `stiff'
combinations of parameters, with an enormous sloppy neutral subspace.
Robustness is often assumed to be an emergent evolved property, but the sloppiness natural to biochemical networks offers an alternative non-adaptive explanation.
Conversely, ideas developed to study evolvability in robust systems can be usefully extended to characterize sloppy systems.
\end{abstract}

\maketitle	


\section*{Introduction}

Robustness and evolvability are major themes of systems biology, have
been the subject of several recent books and
reviews~\cite{Wag05,VisHerWag03,Jen05,LenBarOfr06,FelWag08}, and have
been discussed alongside related phenomena such as canalization,
homeostasis, stability, redundancy, and
plasticity~\cite{Kit07,Jen03,Wag08, KraPlo05}.  Broadly construed,
``robustness is the persistence of an organismal trait under
perturbations''~\cite{FelWag08}, which requires the specification of
both traits of interest and perturbations under consideration.  Recent
work in systems biology has sought to distinguish between 
environmental robustness (e.g., temperature compensation in
circadian rhythms~\cite{Pit54, TomNakKon05, VanLubAlt07}) and 
mutational robustness (e.g., parameter insensitivity in
segment polarity patterning~\cite{DasMeiMun00,ChaSonSen07}). Mutational
robustness has a subtle relation to evolvability; while
allowing survival under genetic alterations, robustness might seem to
reduce the capacity for evolutionary adaptation on multigeneration time
scales~\cite{LenBarOfr06,Wag08}.

Earlier robustness work focused on feedback and control
mechanisms~\cite{BarLei97,AloSurBar99,YiHuaSim00,DoyCse05,Gou04,KurEl-Iwa06}.
Much recent work emphasizes neutral spaces and neutral networks: large
regions in the space of sequences, parameters, or system topologies
that give rise to equivalent (or nearly equivalent) phenotypic
behaviors. Neutral spaces have been explored most extensively in the
context of RNA secondary structure, where large neutral networks of
RNA sequences (genotypes) fold into identical secondary structures
(phenotypes)~\cite{SchFon99, Fon02, SumMarWag07, Wag08}.  More
recently, similar ideas have been applied to neutral spaces underlying
the robustness of gene regulatory networks~\cite{CilMarWag07,
CilMarWag07a,BerSie03}, where different network topologies (genotypes)
can result in identical gene expression patterns (phenotypes). 
Nontrivial niches in sequence spaces are also seen to emerge in
molecular discrimination, a problem where neutral networks allow for
biological communication in the presence of uncertainty akin to that
found in engineered error-correcting codes~\cite{Mye07}.  Functional
redundancies and degeneracies arise at many levels of biological
organization \cite{EdeGal01}, and it is an important open question as
to how neutrality, redundancy, and robustness at different levels are
organized and coupled across scales.

\begin{figure}
\centering
\includegraphics[width=3in]{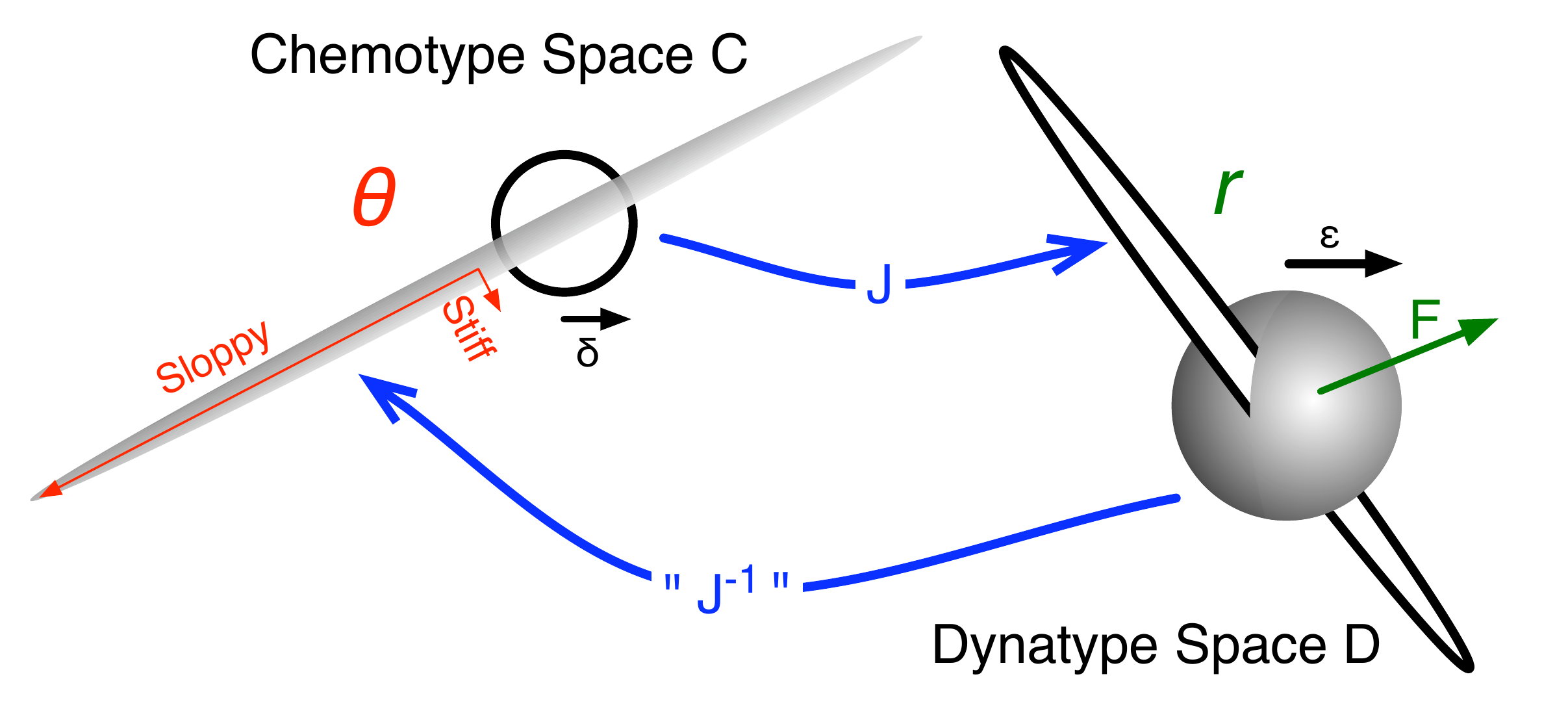}
\caption{\label{fig:Spheres}%
{\bf Sloppiness in the mapping of chemotypes to dynatypes.}
It is natural, at least for cellular regulation and metabolic networks,
to refine the traditional dichotomy of genotype $G$ to phenotype $P$ by
adding two intermediate levels of description, $G \to C \to D \to P$.
Here $C$ is the {\em chemotype}~\cite{GutSet07}, a continuous description
of the behavior in terms of chemical reaction parameters (reaction rates,
barriers and prefactors, or Michaelis-Menten parameters). $D$ is the
{\em dynatype}, meant to describe the dynamical responses of the cell
(usually the time series of all species in response to selected stimuli,
often taken from experimental measurements). Mutations about a particular
chemotype $\btheta$ occupy a region in chemotype space (here a circle of
radius~$\delta$), whose image in dynatype space is given by the local
Jacobian $J$ of the mapping: mutations along stiff directions in chemotype
space will yield large changes in dynatype, while mutations along
sloppy directions will lead
to small dynamical changes. Conversely, a population of individuals
sharing nearly the same dynatype $\br$ (here a sphere of radius~$\epsilon$)
will occupy a distorted region in chemotype space, with large variations
in reaction parameters possible along sloppy directions (gray ellipse).
}
\end{figure}

Despite these advances in understanding neutral networks connecting
genotypes in discrete spaces (e.g., sequences), much of systems
biology is focused on chemical kinetic networks that are parameterized
by continuous parameter spaces.  Often one is interested in the
steady-state behavior of a dynamical system, or in the input-output
response relating only a subset of the chemical
species of a network. In principle, however, one must characterize the full
dynamical behavior of a network, in part because any given
network may be coupled in unknown ways to other subsystems that are not
included in the model.  To more clearly delineate distinct levels of
biological organization, we have chosen to refer the space of
continuous kinetic parameters as a ``chemotype''~\cite{GutSet07}, and
to the full dynamical response of a system as its ``dynatype'' 
(Figure~\ref{fig:Spheres}).  The
chemotype-to-dynatype maps of interest here are embedded within larger
genotype-to-phenotype maps, with chemotypes emerging from lower-level
processes, and dynatypes contributing to phenotypes and ultimately
fitnesses on which selection acts.  Recently, there has been increased
interest in characterizing the parametric sensitivity of the dynamics
of biochemical network models, for two important reasons:
(1) to probe system robustness by quantifying the size and
shape of chemotype spaces that leave system behavior unchanged, and
(2) to characterize system behavior and uncertainties for which
precise values for rate constants and other kinetic parameters are
typically not known.

Parameter estimation in multiparameter models has long been known to
be ill-conditioned: the collective behavior usually cannot be used to
infer the underlying constants.  Recent work has shown that these
models share striking universal
features~\cite{BroSet03,BroHilCal04,GutWatCas07,GutCasWat07}, a
phenomenon that we have labeled ``sloppiness'' (see Figures~\ref{fig:Spheres}
and~\ref{fig:IntersectingSloppiness}).  Sloppiness refers to the
highly anisotropic structure of parameter space, wherein the behavior
of models is highly sensitive to variation along a few `stiff'
directions (combinations of model parameters) and more or less
insensitive to variation along a large number of `sloppy' directions.
A nonlinear least-squares cost function can be constructed:
\begin{equation}
\label{eq:ChrisCostEqn}
C(\btheta)= \sum_i \frac{1}{2} \frac{(x(\btheta)-x_i)^2}{\sigma_i^2}
= \sum_i \frac{1}{2}r_i^2,
\end{equation}
where $r_i=(x(\btheta)-x_i)/\sigma_i$ is the residual describing the
deviation of a dynamical variable $x$ from its measured values $x_i$ with
uncertainty $\sigma_i$. This cost reflects how well
a model with a given set of parameters $\btheta$ fits observed
experimental data.  Parametric sensitivities of the model are encoded
in the Jacobian matrix $J=\partial r_i/\partial \theta_j$.
The curvature of the
cost surface about a best fit set of parameters is described by the
Hessian $H_{mn}=\partial^2 C/\partial \theta_m \theta_n$ (or its 
approximation, the Fisher Information Matrix $J^T J$).
Stiff and sloppy directions are conveniently measured using an analysis of
eigenvalues $\lambda_n$ of the Hessian $H$ (Figure~\ref{fig:EigenvalueFig});
large eigenvalues
correspond to stiff directions.  For a broad range of multiparameter
models (e.g., sixteen models drawn from the systems biology
literature~\cite{GutWatCas07} and models from quantum Monte Carlo,
radioactive decay, and polynomial fitting~\cite{WatCasGut06}) these eigenvalues
are roughly uniformly spread over many decades, with many sloppy
directions a thousand times less well determined than the stiffest,
best constrained parameter combinations.  Two consequences are that
useful model predictions can be made even in the face of huge
remaining parameter uncertainty, and conversely that direct
measurements of the parameters can be inefficient in making more
precise predictions~\cite{GutWatCas07}.  Random matrix theory can be
used to develop insight into the source of this type of eigenvalue
spectrum and the nature of redundancies that appear to underly
sloppiness~\cite{WatCasGut06}.  Our open-source code SloppyCell
(\url{http://sloppycell.sourceforge.net}) provides tools for exploring parameter space
of systems biology models~\cite{MyeGutSet07}.

Others have recently addressed similar questions motivated by the lack
of detailed
information about kinetic parameters.  These include: the inference of
probabilistic statements about network dynamics from probability
distributions on parameter values~\cite{LieKli05}; the use of
``structural kinetic modeling'' to parameterize the Jacobian matrix $J$
and thereby probe ensembles of dynamical
behaviors~\cite{SteGroSel06,GriSelBul07}; the construction of convex
parameter spaces (``k-cones'') containing all allowable combinations
of kinetic parameters for steady-state flux
balance~\cite{FamMahPal05}; the use of ideas from control theory, 
worst-case analysis and hybrid optimization to measure the robustness 
of networks to simultaneous parameter variation~\cite{KimBatPos06},
and exploration of correlated parameter 
uncertainties obtained via global inversion~\cite{PiaFenJia08}.

Can we connect sloppiness to robustness and evolvability?
It is our contention that sloppiness -- the
highly anisotropic structure of neutral variation in the space of
chemotypes -- has important implications for how one characterizes
robustness in systems biology models.  In addition, insights developed
in the study of robustness and evolvability suggest new and
potentially useful ways of analyzing and interpreting sloppiness.

\section*{Environmental robustness and sloppiness}
\label{sec:EnvironmentalRobustness}



\begin{figure}
\centering
\includegraphics[width=3in]{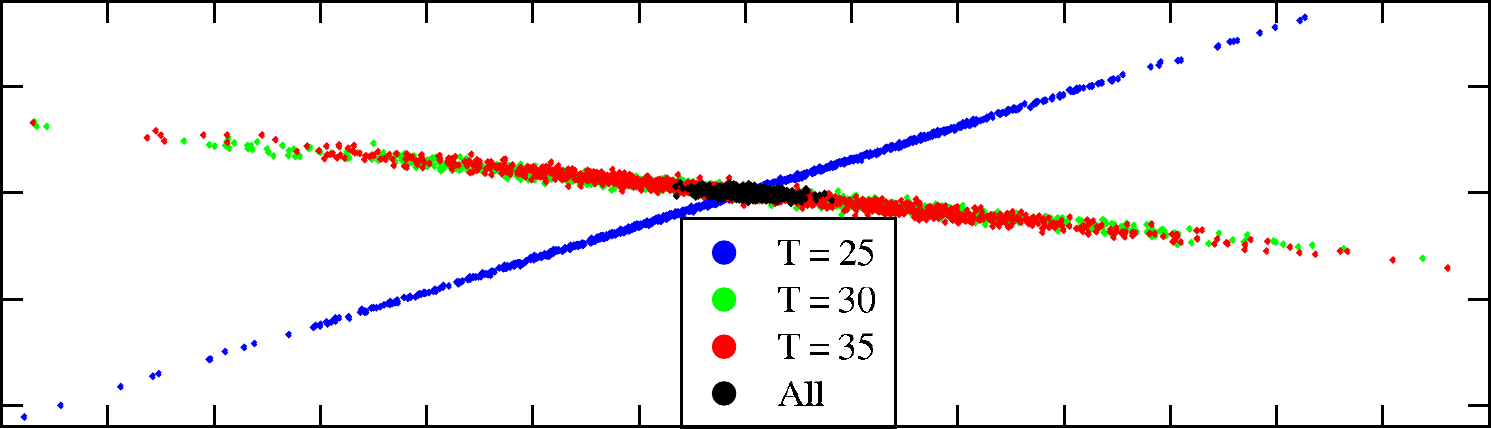}
\caption{\label{fig:IntersectingSloppiness}%
{\bf Sloppy parameter distributions: dependence on external 
conditions.} Shown is a two-dimensional view of the parameter sets
(free energy barriers and prefactors) that 
accurately predict the experimental phosphorylation dynamics~\cite{TomNakKon05}
in a 36-parameter subnetwork of a model of circadian
rhythms~\cite{VanLubAlt07}, within a harmonic approximation (see
Supplemental Material). Shown are parameters valid
at three different temperatures (colors) and valid
for all temperatures simultaneously (black).
The plot shows one `stiff' direction in parameter space for each
temperature which is tightly constrained by the data, and one `sloppy'
direction which has relatively large variations without change in behavior.
Most of the 34 other directions in parameter space not shown are sloppy;
the two-dimensional view was chosen to best align with the stiffest
direction for each of the four ensembles. The black region models
organisms that are robust to temperature changes in this range.
The acceptable region rotates and shifts with temperature, but the 
sloppiness allows different temperatures to intersect (robust temperature
compensation) even though all rates are strongly temperature dependent.
}
\end{figure}

\begin{figure}
\centering
\includegraphics[width=3in]{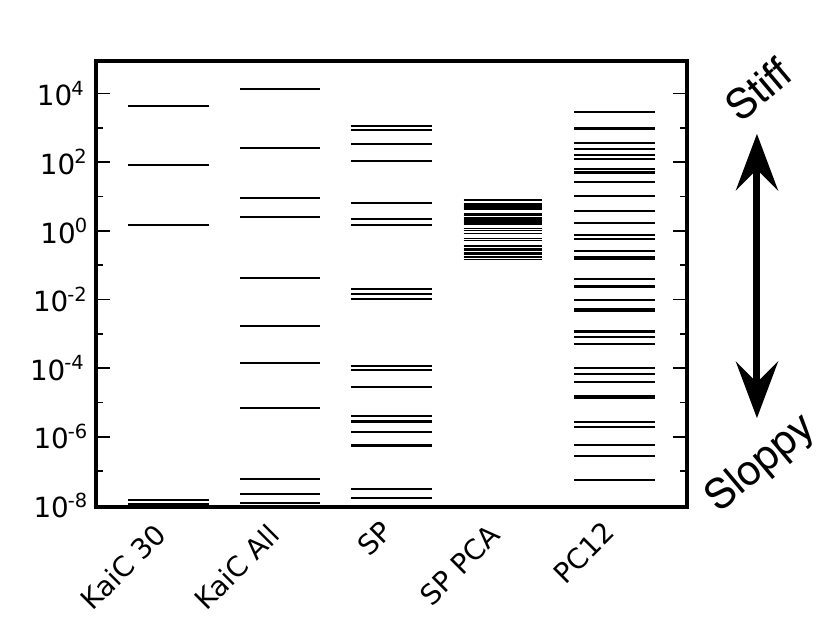}
\caption{\label{fig:EigenvalueFig}%
{\bf Sloppy model eigenvalues.} Shown are the eigenvalues of the 
approximate Hessian $J^T J$ for the goodness-of-fit $C(\btheta)$ 
(Equation~\ref{eq:ChrisCostEqn}) about the best fit. Large eigenvalues
correspond to stiff directions; others are sloppy. Notice
the enormous range on this logarithmic scale; not all eigenvalues 
(ranging down to $10^{-20}$) are depicted. 
\hfill\break$\bullet$
Columns {\em KaiC 30} and 
{\em KaiC All}
are for the KaiC phosphorylation dynamics model 
(Figure~\ref{fig:EigenvalueFig}), showing
$T=30^\circ$ C (yellow region in Figure~\ref{fig:IntersectingSloppiness})
and simultaneous fits for all temperatures (black region). Notice that the
`robust' simultaneous fit has roughly one more stiff direction than the 
single temperatures. 
\hfill\break$\bullet$
The {\em SP} and {\em SP PCA} columns are for the segment 
polarity model~\cite{DasMeiMun00,MeiMunOde02}.
{\em SP} is an eigenvalue analysis about one of the 
acceptable parameter sets, showing parameters that keep the behavior
(dynatype) of the entire network preserved (time series for all components
under all experimental conditions). {\em SP PCA} is a principal components
analysis of the segment polarity ensemble that yields the wild-type
phenotype, with parameters restricted to a relatively small range (roughly
three decades each). Most directions in {\em SP} are sloppy
enough to have fluctuations larger than the sampled phenotype box in 
{\em SP PCA}; the sloppy dynatype {\em SP} already explains the robustness
to all but a few stiff directions in parameter space. Conversely,
the sensitivity of the dynatype {\em SP} to a few stiff directions does
not preclude phenotypic robustness in those directions for {\em SP PCA};
the dynatype (all dynamical evolution) is far more restrictive than the
phenotype (output patterning).
\hfill\break$\bullet$
{\em PC12} is for the EGF/NGF growth-factor signaling
network~\cite{BroHilCal04,GutWatCas07}; note that it too is sloppy. See
Figure~\ref{fig:Evolvability} for an analysis of evolvability and robustness
for this model.
}

\end{figure}

Organisms must thrive under many environmental conditions: changing
temperatures, salt concentrations, pH, nutrient densities, etc.  Many
organisms have explicit control mechanisms to keep their internal
state insensitive to these external changes -- these control
mechanisms (homeostasis, adaptation, etc.) have been a historical
focus in the robustness literature \cite{BarLei97,SteSauSza04}. 
For variations in temperature, however, many organisms do not have
such homeostatic control (with the exception of birds, mammals, and
some plants) and must instead cope with the exponential Arrhenius 
temperature dependence of all their reaction rates by some sort of 
compensatory mechanism~\cite{RuoRenKom97}.

The prototypical example of temperature compensation is the 24-hour
period of circadian rhythms~\cite{Pit54}.  Recent experiments have
succeeded in replicating the circadian control network of
cyanobacteria in the test tube using three Kai proteins, whose degree
of phosphorylation oscillates with a temperature-compensated period in the
range of 25 to 35$^\circ$ C.  In addition, the
phosphorylation dynamics of KaiC alone is found to be unchanged as the
temperature varies in the same range~\cite{TomNakKon05}. This has been
cited as a plausible explanation for the observed temperature
compensation in the full network, presuming that 
all other rates are fast~\cite{VanLubAlt07} and hence irrelevant to the
period. (At least one other
explanation of temperature compensation~\cite{HonConTys07} also relies
on constraining most
rates to be irrelevant). Narrowing our focus to the KaiC phosphorylation
subnetwork, however, still leaves the nontrivial task of explaining its
temperature compensation mechanism, 
since estimated energy barriers~\cite{CheZhaMcC06} suggest that phosphorylation 
rates should be twice as fast at the higher temperature.

The dynamics of KaiC phosphorylation have been modeled using
six phosphorylation sites and two conformational
states (active and inactive)~\cite{VanLubAlt07}.  If each of the 18
rates in this model roughly double between 25 and 35$^\circ$C,
can we adjust the corresponding energy barriers and prefactors such
that the resulting net phosphorylation dynamics is
temperature-independent?

Figure~\ref{fig:IntersectingSloppiness} shows a two-dimensional view
of the acceptable parameter sets in the resulting 36-dimensional space of
energy barriers and prefactors, 
explored in the harmonic approximation (see Supplemental Material).
Notice that the region of
acceptable parameters rotates and shifts as the temperature changes.
Notice also that the system is sloppy:
Figure~\ref{fig:IntersectingSloppiness} shows one stiff direction that
is highly constrained by the data and one sloppy direction that is
largely unconstrained.  The eigenvalue analysis in
Figure~\ref{fig:EigenvalueFig} confirms that most directions in
parameter space are sloppy and unconstrained.  This provides a natural
explanation for robustness: the intersection of these large, flat
hypersurfaces yields parameters that work at all temperatures.%
  \footnote{In the particular case of KaiC, we find that successful 
  chemotypes favor dephosphorylation in the active state and 
  phosphorylation in the inactive state  (see
  Supplemental Material), so
  the thermally robust solutions presumably increase the proportion of
  protein in the inactive state as temperature increases, compensating 
  for the general speedup of all rates.}
In general,
each external condition provides one constraint per stiff direction;
since there are only a few stiff directions and many parameters in
sloppy models, robust behavior under varying external conditions is
easily arranged. Indeed, Figure~\ref{fig:EigenvalueFig} shows that the
robust, temperature-independent fits for the KaiC model are themselves
a sloppy system.

\section*{Chemotype robustness and sloppiness}

In addition to robustness to environmental perturbation, biological
networks are often robust to mutational perturbations;
they maintain their function in the face of mutations that
change one or perhaps more of their underlying rate parameters, and
thus change their location in chemotype space.
Some authors have used
this as a criterion for judging model plausibility~\cite{MaLaiOuy06}.
The quintessential example of a system that is chemotypically robust
is the \emph{Drosophila} segment polarity gene network.  Early in
development, this network generates a periodic macroscopic 
phenotype: a pattern of gene
expression across several cells that persists throughout development
and guides later stages. Multiparameter models of this
network~\cite{DasMeiMun00,ChaSonSen07,MaLaiOuy06,Ing04} find that 
a surprisingly large fraction of randomly chose parameter sets
generate a pattern consistent with the observed patterning of three
genes -- the system exhibits chemotype robustness.


In the context of sloppy models, we may define chemotype
robustness as the fraction of a given volume in parameter/chemotype space $C$
that maps into a functional region of behavior/dynatype space $D$ 
(Figure~\ref{fig:Spheres}). This latter functional region 
represents behavior close to optimum (or close to that measured 
experimentally). For simplicity, let us consider it 
to be a hypersphere of radius~$\epsilon$ (i.e., a cost 
$C(\btheta) = \sum{r_i^2}/2 < \epsilon^2/2$ in Equation~\ref{eq:ChrisCostEqn});
larger changes in behavior are considered significantly different,
perhaps lowering the organism's fitness. The given volume in chemotype
space $C$ might be (as for the segment polarity network) a hypercube
of parameter ranges deemed reasonable, or (as a simple model of mutations)
a hypersphere; let its scale be given by $\delta$. Our robustness
is therefore the fraction of all points in the $\delta$-ball in $C$
that map into the $\epsilon$-ball in $D$ -- in Figure~\ref{fig:Spheres}
the fraction of the circle whose interior is colored gray. This fraction
can be calculated (see Supplemental Material) and is approximately given
by 
\begin{equation}
\label{eq:ChemotypeRobustness}
R_c = \prod_{\lambda_n>\lambda_{crit}} \sqrt{\frac{\lambda_{crit}}{\lambda_n}},
\end{equation}
where $\lambda_{crit}=\epsilon^2/\delta^2$. This formula can be motivated
by considering the robust subregion (gray needle intersecting the circle)
to be a slab, with thickness $\epsilon\sqrt{\lambda_n}$ along the 
eigendirection corresponding to each eigenvalue $\lambda_n$.%
  \footnote{The cost for a small displacement of size $\Delta \theta$ along
  the eigendirection $n$ is $\lambda_n \Delta\theta^2/2$, which equals
  $\epsilon^2/2$ when $\Delta \theta = \pm \epsilon \sqrt{\lambda_n}$.}
For sloppy directions with $\lambda_n < \epsilon^2/\delta^2 = \lambda_{crit}$,
the slab is thicker than the circle and does not reduce the robust fraction;
for each stiff direction with $\lambda_n > \lambda_{crit}$, the fractional
volume is reduced roughly by a factor of the slab thickness 
$\epsilon \sqrt{\lambda_n}$ over the sphere width $\delta$, leading to
Equation~(\ref{eq:ChemotypeRobustness}).

In their model of segment polarity, von Dassow et al.\ found that
approximately one in 200 randomly chosen parameter sets generated a
wild-type expression pattern for three key genes~\cite{DasMeiMun00}. 
This would naively
seem amazing for a 
48 parameter model like theirs; in an isotropic approximation, each parameter
would be allowed only 6\% chance of changing the wild-type pattern
(since $0.94^{48} \sim 1/200)$. However, we have previously shown
that the segment polarity model is
sloppy~\cite{GutWatCas07}. That is, going far beyond restricting
the output phenotype, the dynamical evolution of every component
of the network is approximately
preserved even with huge changes in parameter values: only a few 
stiff directions in chemotype space are needed to maintain the
dynatype (see column {\em SP} in Figure~\ref{fig:EigenvalueFig}). 
Sloppiness hence provides a natural explanation for the wide variations
in all but a few directions in parameter space. 

The success rate of
one in 200 is not nearly as striking if the dynamics is already known
to be insensitive to all but perhaps four or five combinations of parameters:
$0.35^5 \times 1^{43} \sim 1/200$. Column {\em SP PCA} in 
Figure~\ref{fig:EigenvalueFig} fleshes this picture out with a principal
components analysis (PCA) of the robust region seen in von Dassow et al.'s original
model, reconstructed using Ingeneue~\cite{MeiMunOde02}. Note that
these PCA eigenvalues are cut off from below by the parameter ranges
chosen by the original authors for exploration (typically three decades
per parameter).
While the overall scale of the dynatype sloppy-model eigenvalues in 
{\em SP} and the phenotype eigenvalues in {\em SP PCA} cannot be directly
compared, it is clear that the vast majority of sloppy-model eigenvalues are 
too small to constrain the parameters within the explored region. The
model is robust in these directions not because of evolution and fitness,
but because the dynamics of chemical reaction networks is mathematically
naturally dependent only on a few combinations of reaction parameters.


\section*{Robustness, evolvability, and sloppiness}

\begin{figure}
\centering
\includegraphics[width=3in]{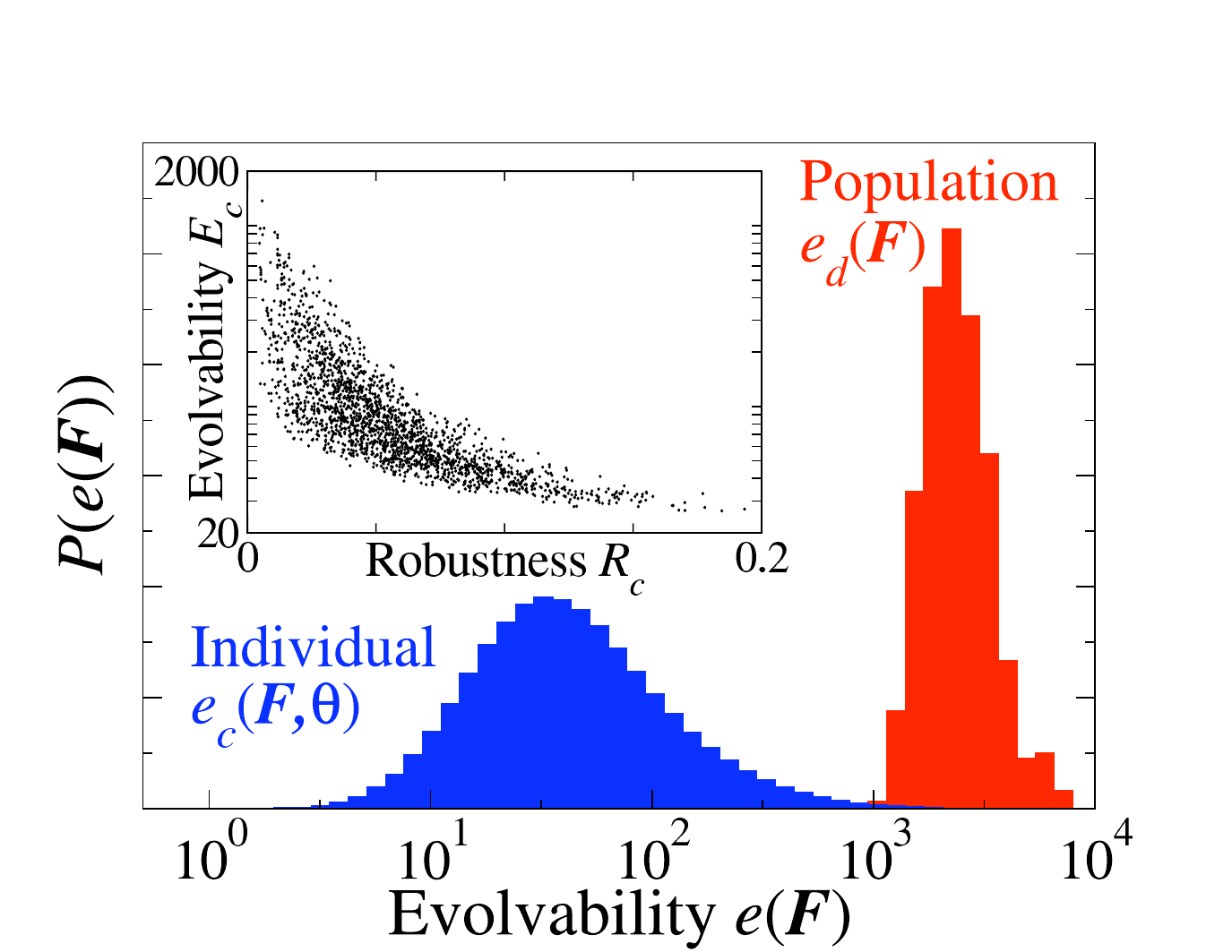}
\caption{\label{fig:Evolvability}%
{\bf Evolvability and robustness in a sloppy system.}
Evolvability distributions, and evolvability versus robustness, for an
ensemble of parameters for a model of an EGF/NGF signaling pathway
fitted to experimental data in PC12 cells~\cite{GutWatCas07}.
The histogram on the left is the distribution of 
individual/chemotype evolvabilities $e_c(\bF,\btheta_\alpha)$
(Equation~\ref{eq:ChemotypeEvolvability}), as $\bF$
(an evolutionary pressure in dynatype space) is randomly chosen in direction
with uniform magnitude and $\btheta_\alpha$ varies over the ensemble. The
histogram on the right is the corresponding
distribution of population/dynatype evolvabilities 
$e_d(\bF)$ (Equation~\ref{eq:DynatypeEvolvability}). Note that the 
population evolvabilities are significantly higher than the individual
ones. The inset plots the RMS individual chemotype evolvability
$E_c(\btheta_\alpha)$ versus the robustness $R_c(\btheta_\alpha)$
(Equation~\ref{eq:ChemotypeRobustness}) for the ensemble. 
($\lambda_{crit}$ is chosen as the fourth-stiffest eigenvalue
at the best fit: see Supplemental Material). Note that,
for each individual, more robustness leads to less evolvability -- 
individuals which rarely mutate to new forms can't evolve as readily.
This need not apply to the population, insofar as we expect robust
dynatypes to explore larger regions of parameter/chemotype space,
and thus the ratio of dynatype to chemotype evolvability to increase
with increasing robustness.}
\end{figure}

Mutational robustness of systems would seem to be at odds with
an ability to adapt and evolve, since robustness implies persistence
of phenotype or function, which may inhibit the capacity for
evolutionary change.  The concept of neutral spaces has been used --
most notably by Wagner and collaborators -- to suggest a resolution of
this apparent paradox, as demonstrated in model systems exploring
various genotype-to-phenotype maps~\cite{Wag08, SumMarWag07, CilMarWag07, CilMarWag07a}.  The important insight
is that neutral spaces and neutral networks enable systems to drift
robustly in genotype space (i.e., without significant phenotypic
change), while encountering new and different phenotypes at various
points along that neutral space.  This insight results from a distinction
between the robustness and evolvability of any given genotype, and 
the robustness and evolvability of all genotypes consistent with a 
given phenotype~\cite{Wag08}.

Evolvability is postulated to reflect the range of possible different 
phenotypes that are possible under genotypic mutation.
How does the sloppy connection between parameters and behavior impinge on
the question of evolvability?  Translating previous work on discrete 
genotype and phenotype spaces to the continuous spaces of chemotypes 
and dynatypes is nontrivial.  Since the dimensionality of 
the space of chemotypes is less than that of dynatypes, the volume 
of dynatype space accessible under changes in chemotype is zero, 
i.e., lies on a lower-dimensional subspace.  To develop a sensible 
definition of evolvability in such systems, we postulate forces $\bF$ 
in dynatype space (Figure~\ref{fig:Spheres}) that reflect evolutionary
pressures due to changes 
in the environment, such that a change $\br$ in dynatype leads to 
a change $\br \cdot \bF$ in fitness.  An organism's evolvability is
related to its 
capacity to respond to external forces through appropriate mutations
in chemotype. 

For a given force $\bF$, the maximum fitness change among 
mutations of size $\delta$ in chemotype space is given by:
\begin{equation}
\label{eq:ChemotypeEvolvability}
e_c(\bF, \btheta) = \sqrt{\bF^{T} J J^{T} \bF}\, \delta
\end{equation}
which we call the chemotype evolvability distribution (see Supplemental
Material).
Refs.~\cite{BroHilCal04} and~\cite{GutWatCas07} generate ensembles of
parameters (chemotypes)
consistent with a given dynatype for an EGF/NGF signaling pathway in
PC12 cells, where the dynatype is constrained to fit available
experimental data. (The PC12 network is sloppy, see
Figure~\ref{fig:EigenvalueFig}.)
Each member of such an ensemble $\btheta_\alpha$ has a
Jacobian $J_\alpha$.  As in Ref.~\cite{Wag08}, which distinguishes between
genotype and phenotype evolvability, we can distinguish between 
the chemotype $e_c(\bF,\btheta_\alpha)$ and 
dynatype 
\begin{equation}
\label{eq:DynatypeEvolvability}
e_d(\bF) = \max_{\theta_\alpha} e_c(\bF,\btheta_\alpha)
\end{equation}
evolvability distributions.  The first
gives the distribution of adaptive responses to $\bF$
of individual chemotypes in a population, while
the second gives the optimal response within the
population.  Figure~\ref{fig:Evolvability} shows the chemotype and 
dynatype evolvability distributions, generated using
the PC12 ensemble of Ref.~\cite{GutWatCas07} and a
uniform distribution of force directions $\bF$ in dynatype
space.  Within a population sharing the same behavior, we find
substantial variation of accessible behavior changes, leading to a
substantially larger population (dynatype) evolvability than individual
(chemotype) evolvability.  This echoes the finding of Wagner that phenotype
evolvability is greater than genotype evolvability for RNA secondary
structures~\cite{Wag08}.

It is natural to define an overall evolvability as the root-mean-square
average of the evolvability distribution over a spherical distribution
of environmental forces $\bF$ in dynatype space:
\begin{equation}
\label{eq:Ec}
E_c(\btheta_\alpha) = \sqrt{\langle (e_c(\bF,\btheta_\alpha)^2 \rangle}_\bF
\end{equation}
and correspondingly for the overall RMS dynatype evolvability. The
inset to Figure~\ref{fig:Evolvability} shows that the chemotype evolvability
decreases as the chemotype robustness increases,
closely analogous to Wagner's discovery that genotype evolvability decreases
as genotype robustness increases, except that his plot averages over
phenotypes while ours represents variation within a dynatype. Thus we
reproduce Wagner's observation~\cite{Wag08} that individual evolvability
decreases with robustness and that population evolvability is significantly
larger than individual evolvability.%
  \footnote{Unfortunately, we cannot reproduce Wagner's final conclusion
  (that phenotype evolvability increases with phenotype robustness), since our
  ensemble (generated to match experimental behavior) is confined to 
  the single PC12 species (dynatype).}

\section*{Conclusion}

Our previous work aimed at developing predictive systems biology models 
in the face of parametric uncertainty has led us to formulate a theory 
of sloppiness in multiparameter models.  The picture that emerges from 
this theory is of a highly anisotropic neutral space in which variation 
in parameters (chemotypes) can leave system behavior (dynatypes) unchanged.
This picture is reminiscent in many ways to the notion of neutral spaces
and neutral networks that has been developed to explore the robustness 
and evolvability of biological systems.  We have been motivated by those 
ideas to here reconsider sloppiness within that context, both to highlight 
implications of sloppiness for the study of robustness and evolvability, 
and to identify new methods for analyzing sloppy systems.

\section*{Acknowledgments}
We would like to thank Ben Machta and Mark Transtrum for their keen
insights into sloppiness and their assistance in developing some of
the arguments presented here.  We acknowledge grants USDA-ARS
1907-21000-027-03, NSF DMR-070167, and NSF DGE-0333366.



\bibliography{SloppyRobust}

\clearpage


\begin{center}
\Large{Supplemental Material} \\ ~ \\ \large{\emph{Sloppiness, robustness, and evolvability in systems biology}}
\end{center}
\subsection*{Bryan C. Daniels, Yan-Jiun Chen, James P. Sethna, 
Ryan N. Gutenkunst, and Christopher R. Myers \\ ~ \\}

\renewcommand{\thepage}{S\arabic{page}}
\setcounter{page}{1}
\renewcommand{\theequation}{S\arabic{equation}}
\setcounter{equation}{0}
\renewcommand{\thefigure}{S\arabic{figure}}
\setcounter{figure}{0}

\section*{Contents}

The contents of the supplemental material are organized corresponding to the order of the main text.
Included in the supplemental material are derivations of mathematical results and details of the specific models mentioned
in the main text.  

We have also posted the data files and computer codes for the models discussed, at \url{http://www.lassp.cornell.edu/sethna/Sloppy}.
For the KaiC, PC12, and segment polarity models, this includes:
\begin{enumerate}
\item Equations in \LaTeX, Python, and C
\item SBML (system biology markup language) files
\item Parameter ensembles
\item Best-fit Hessian and $J^T J$, and their eigenvectors and eigenvalues
\end{enumerate}

\section*{Introduction}
\begin{center}
\textit{Hessian at best fit parameters}
\end{center}

In the introduction we mention that ``the curvature of the cost surface about a best
fit set of parameters is described by the Hessian $H_{mn}$.''  Examining the behavior of $H_{mn}$ 
is a standard method for nonlinear least squares models when fitting data.  Formally, $H_{mn}$ is written as:   
\begin{equation}
H_{mn} 	= \frac{\partial^2 C}{\partial \theta_m \theta_n} 		
	= \sum_i \frac{\partial r_i}{\partial \theta_m}
		  \frac{\partial r_i}{\partial \theta_n}
	    + r_i \frac{\partial^2 r_i}{\partial \theta_m \theta_n}.
\end{equation}
If the model fits the data well so that $r_i \approx 0$ (or perfectly,
Ref.~\cite{GutWatCas07} in the main text) then
\begin{equation}
\label{eq:HJTJ}
H_{mn}(\btheta^*)\approx\sum_i \frac{\partial r_i}{\partial \theta_m} 	
			 \frac{\partial r_i}{\partial \theta_n}	
	= (J^T J)_{mn}.
\end{equation}
If $H$ and the cost are used (as in this review) to describe changes in model
behavior from $\btheta^*$, then $\br \equiv 0$ at $\btheta^*$ and 
Equation~(\ref{eq:HJTJ}) is exact.
Notice also that $H$ reflects the sensitivity of the fit to changes in parameters; in fact, its inverse is the covariance matrix. The diagonal elements of the covariance matrix are proportional to the uncertainties in the parameters, while the off-diagonal
elements are estimates of parameter uncertainty correlations.

\begin{center}
\textit{Figure \ref{fig:Spheres}: Sloppiness in 
the mapping of chemotypes to dynatypes}
\end{center}

Shown in Figure \ref{fig:Spheres} of the main text is the mapping of
chemotypes $C$ to dynatypes $D$.  The
mapping between $C$ and $D$ is described with $J$ and ``$J^{-1}$''.  It is
typical that $\dim(C) \ll \dim(D)$, since there are typically more data
points constraining the dynatype than there are parameters defining
a chemotype.  Therefore, the inverse of $J$ is not well-defined.  
In Figure \ref{fig:Spheres}, the gray ellipse in $C$ represents the 
inverse image of the $\epsilon$-ball, $B_{\epsilon}$, in $D$ under $J$. 
That is, ``$J^{-1}$'' acting on $B_{\epsilon}$ is the set 
$\{ \mathbf{c} \in C~s.t.~J \cdot \mathbf{c} \in B_\epsilon \}$.  

Note also that the stiff and sloppy eigendirections in $C$ and their images
in $D$ can be described by the singular value decomposition of the
Jacobian $J$. Since $\lambda_n$ are eigenvalues of 
$J^T J$, $\sqrt{\lambda_n}$ are the singular values of $J$. Furthermore, writing $J = U \sum V^T$, we see that the columns of $V$ are stiff/sloppy eigenparameters in $C$ (shown in red in the figure), and the columns of $U$ are images of stiff and sloppy eigenparameters (divided by $\lambda_n$) in $D$.



\section*{Environmental robustness and sloppiness}

\begin{center}
\textit{Figure \ref{fig:IntersectingSloppiness}: Sloppy parameter
distributions: dependence on external conditions}
\end{center}

In Figure~\ref{fig:IntersectingSloppiness} of the main text, the plane
onto which the ensembles are projected is the one
that aligns best with the stiffest eigenparameter of each of the four 
ensembles.  
To accomplish this, the vertical and horizontal axes in
Figure~\ref{fig:IntersectingSloppiness} are, respectively, the first and second singular vectors
in the singular value decomposition of the set of stiffest eigenparameters
$\{\mathbf{v}_0^{25},\mathbf{v}_0^{30},\mathbf{v}_0^{35},\mathbf{v}_0^{All}\}$.
In a way analogous to principal components analysis, this gives the plane
that passes through the origin and comes closest to passing through
the heads of unit vectors pointing in the stiffest eigendirections.

Each ensemble of parameter sets shown in 
Figure~\ref{fig:IntersectingSloppiness} is chosen from the
probability distribution corresponding to the local quadratic approximation 
of the cost near the best-fit parameters $\btheta^*$:
\begin{equation}
P( \btheta^* + \Delta \btheta ) \propto \exp( -\Delta \btheta J^TJ \Delta \btheta /2).
\end{equation}
This local approximation to the cost was used to generate the
ensembles instead of the full nonlinear cost function due to difficulties
in generating equilibrated ensembles: the thin curving manifolds of allowable
chemotypes for sloppy models can be notoriously difficult to populate.  But
this is not impossible; efforts are still underway, and if equilibrated
ensembles are found, they will be posted to the website mentioned above.

\begin{center}
\textit{KaiC phosphorylation subnetwork model}
\end{center}

\begin{figure}
\centering
\includegraphics[width=3in]{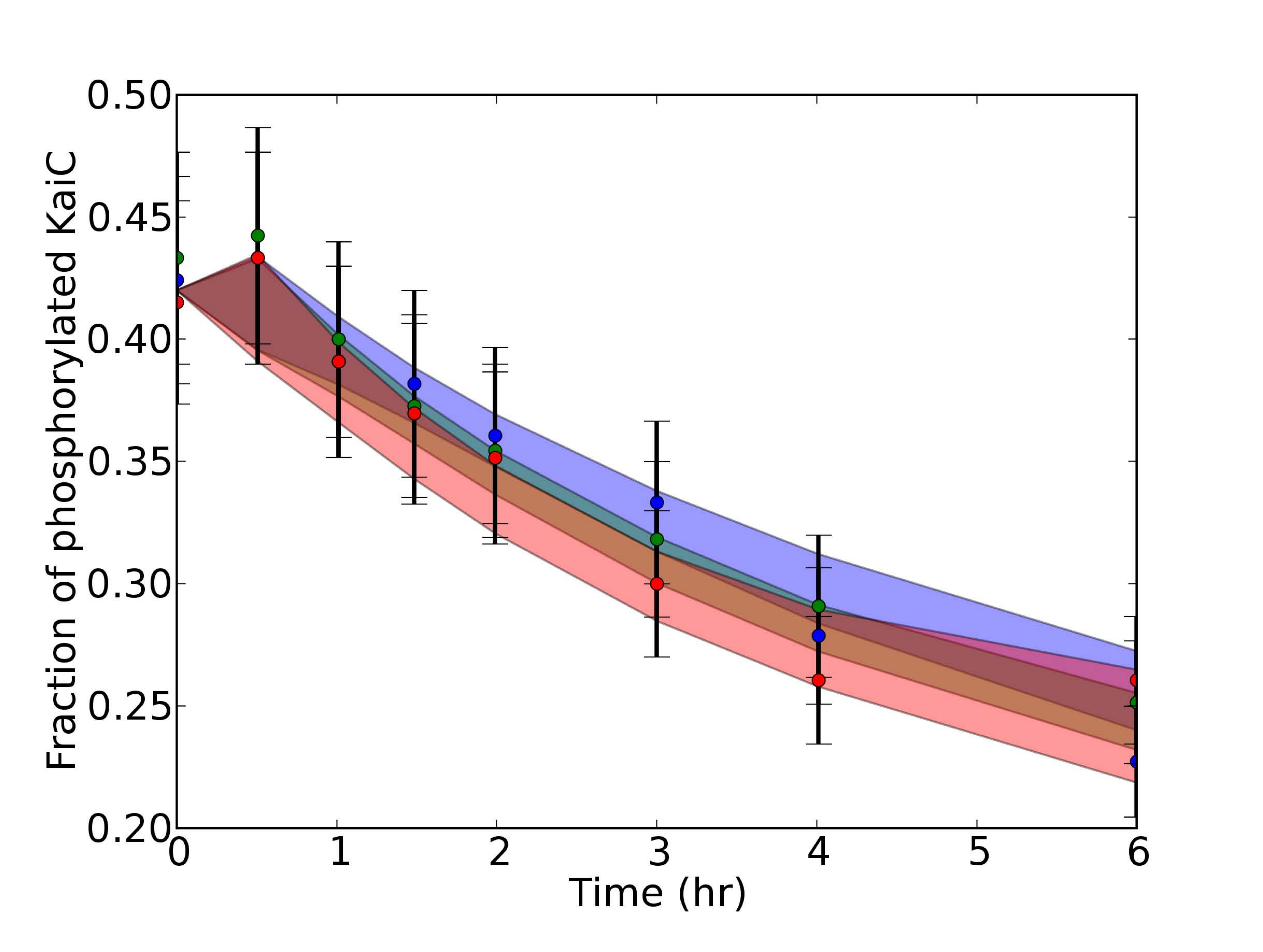}
\caption{\label{fig:EnsembleOutput}%
{\bf KaiC phosphorylation network: temperature-compensated output.}
Shown is the net phosphorylation of KaiC over time, comparing 
experimental data (circles with error bars, from Ref.~\cite{TomNakKon05}
in the main text) 
with output from an ensemble of chemotypes (filled colored regions, showing
the mean plus or minus one standard deviation over the ensemble for the
net phosphorylation at each time-point).
Different colors correspond to different temperatures: 
blue = 25$^\circ$, green = 30$^\circ$,
red = 35$^\circ$.  Note that the chemotypes describe the data well at all
three temperatures, even though the rates are strongly dependent on
temperature.
}
\end{figure}

In the main text, 
we use as an example a portion of the circadian
rhythm model presented in Ref.~\cite{VanLubAlt07} of the main text.
We implement the subnetwork that van Zon et al.\ hypothesize
must have intrinsically temperature-independent rates: that which
controls the phosphorylation
of KaiC alone.  This subnetwork models the experimental measurements of KaiC phosphorylation
in the absence of KaiA and KaiB (Ref.~\cite{TomNakKon05} of the main text),
in which the phosphorylation of KaiC does not
oscillate, but decays at a temperature-compensated rate
in the range from 25 to 35$^\circ$~C (see circles in 
Figure~\ref{fig:EnsembleOutput}).

\begin{figure}
\centering
\includegraphics[width=3in]{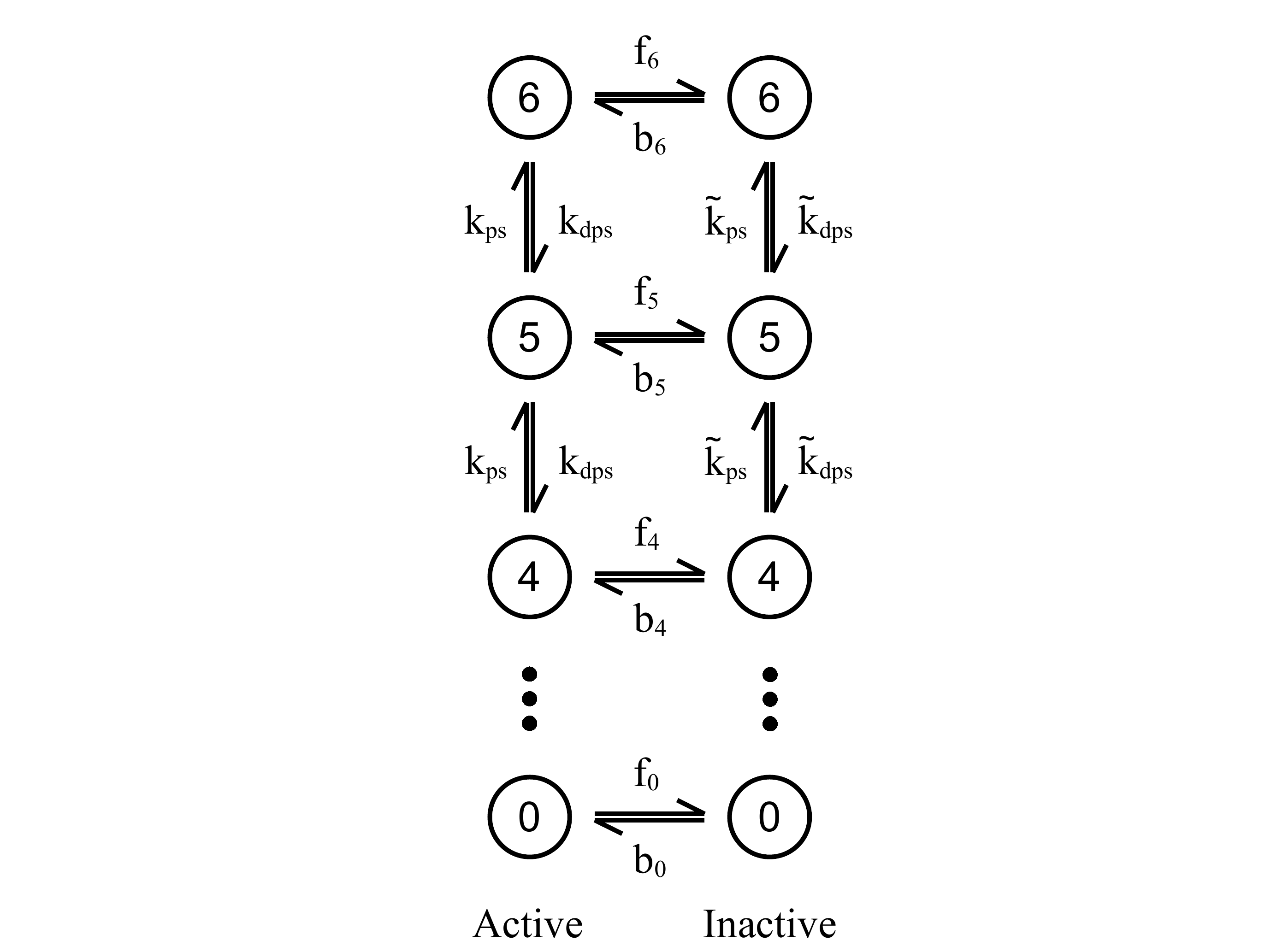}
\caption{\label{fig:KaiCNetwork}%
{\bf KaiC phosphorylation subnetwork.}
This schematic depicts the KaiC network used as an example in the main
text.  It is a portion of the full circadian rhythm model presented in
Ref.~\cite{VanLubAlt07} of the main text.  The numbers represent the degree of phosphorylation,
and the two columns represent two different conformational states, ``active''
and ``inactive.''   The labels on the arrows represent reaction rates for
changing among the phosphorylation and conformation states.  Each 
conformation state has one phosphorylation and one dephosphorylation rate,
independent of the degree of phosphorylation.  Each of the 14 ``flip'' rates
between conformational states ($b_i$ and $f_i$) is allowed to vary 
independently.  This gives a total of 18 reaction rates.
}
\end{figure}

The subnetwork involves an active and inactive state of
KaiC, along with six phosphorylation sites for each state, as depicted
in Figure~\ref{fig:KaiCNetwork}.  Including forward and backward ``flip''
rates between active and inactive states along with (de)phosphorylation rates
that are each constant for the two states, there are 18 independent rates.  
To assess the temperature dependence, we assume that each transition rate
follows an Arrhenius law, with constant energy barrier $E$ and 
prefactor $\alpha$: the $i$th rate is $\alpha_i e^{E_i/kT}$.  This then
gives a 36-dimensional chemotype space in which to search for solutions.

Temperature-independent solutions can be trivially found in this space
if the energy barriers are chosen to be small, since this produces rates
that are inherently weakly dependent on temperature.
In order to avoid this trivial temperature compensation, 
we apply a prior that favors solutions with phosphorylation 
energy barriers near the expected $E_0=23~kT$, similar to those found in other kinases (Ref.~\cite{CheZhaMcC06} in the main text) and appropriate for
reactions that break covalent
bonds. We choose this prior as a quartic in $\log E$:
\begin{equation}
C_{prior} = \frac{25}{2} \left[\log\left(\frac{E}{E_0}\right)\right]^4.
\end{equation}
The form was chosen to severely penalize barriers less than $10~k T$,  but to
be reasonably flat around $E_0$; other prior choices would presumably
perform similarly.

Using this method, we find that it is possible to fit
the experimental data even with (de)phosphorylation
rates that are strongly temperature-dependent. The 
phosphorylation and dephosphorylation rates that provided a best fit
to all temperatures simultaneously were all above 21 $kT$.
We used Bayesian Monte-Carlo sampling of chemotype space to
create an ensemble of parameter sets that each  
produce phosphorylation dynamics that match the
experimental data at 25, 30, and 35$^\circ$~C.  As
explained above, our ensemble has not yet sampled all the space available,
but we still find many such acceptable chemotypes.
The minimum (de)phosphorylation rate for the ensemble was just under 10 $kT$,
so the prior worked as designed to confine the barriers to physically
reasonable values.
Figure~\ref{fig:EnsembleOutput} shows the output of the model over this
ensemble of parameter sets compared with the experimental data
from Ref.~\cite{TomNakKon05} of the main text.  

\begin{figure}
\centering
\includegraphics[width=3in]{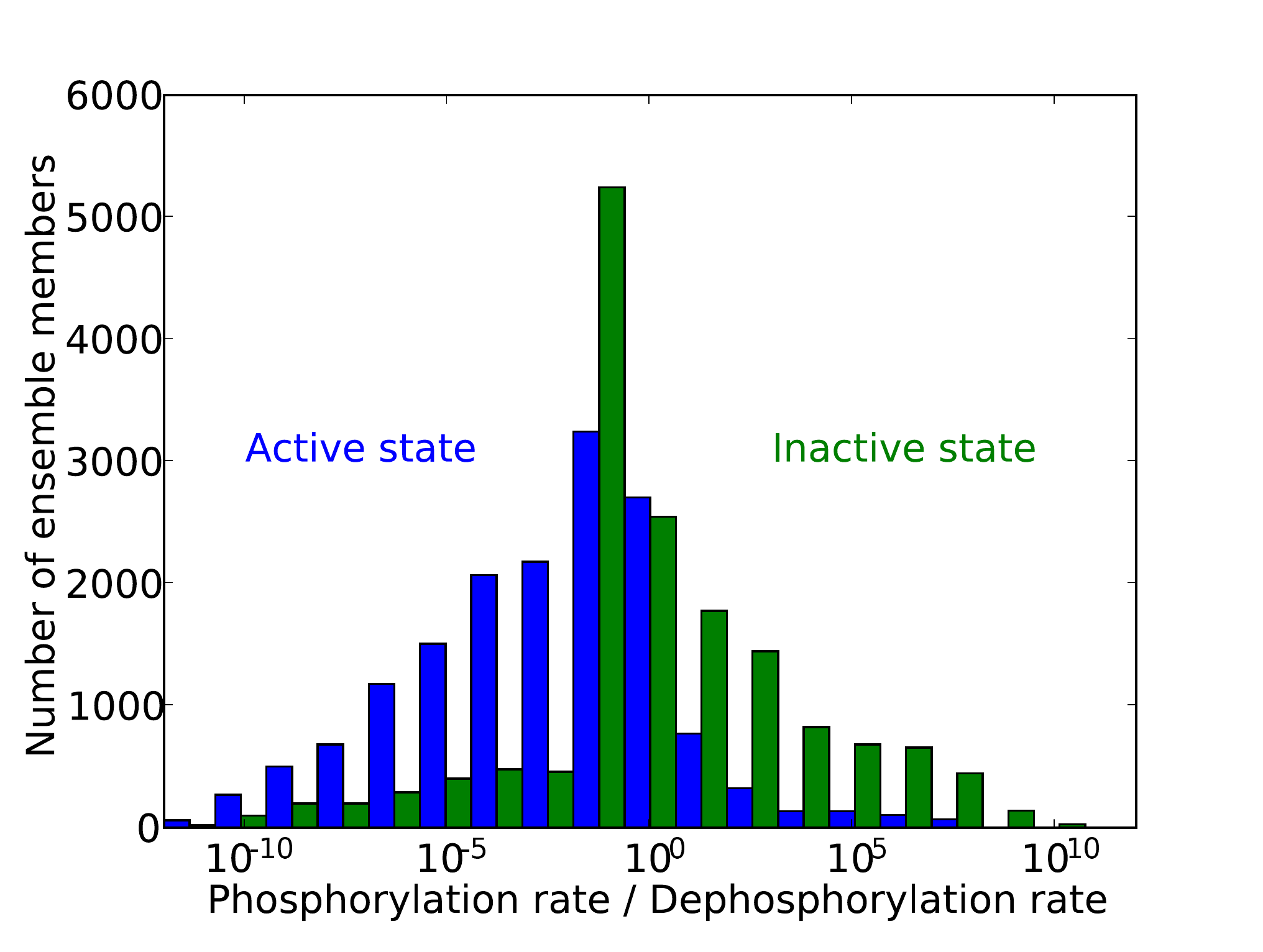}
\caption{\label{fig:PhosRatios}%
{\bf KaiC phosphorylation network: temperature-compensation mechanism.}
This plot shows the ratios of phosphorylation rates to dephosphorylation
rates for the active and inactive states -- the distribution of
k$_{ps}$/k$_{dps}$ is shown in
blue for the active state, and the distribution of \~k$_{ps}/$\~k$_{dps}$
is shown in green for the inactive state (see Figure~\ref{fig:KaiCNetwork}
for definitions of rate constants).  
The distribution is over the 
same (non-equilibrated) ensemble as was used to generate 
Figure~\ref{fig:EnsembleOutput}.  Note that phosphorylation is favored in
the inactive state, while dephosphorylation is favored in the active state.
This suggests a temperature-compensation mechanism, as described in the text.
}
\end{figure}

We mention in a footnote that, in our model, 
``successful chemotypes favor dephosphorylation 
in the active state and phosphorylation in the inactive state.''  This
can be seen in the ratio of phosphorylation to dephosphorylation
rates, shown in Figure~\ref{fig:PhosRatios}, for 
the ensemble of successful chemotypes.
Note that most members of the ensemble have an inactive state with
higher phosphorylation rate than dephosphorylation, and vice versa for
the active state.  This matches with an intuitive temperature-compensation
mechanism: with flip rates that are also temperature-dependent, higher
temperatures can lead to more KaiC being in the inactive state, leading
to a slower overall decay in phosphorylation that compensates for the
speedup in reaction rates.

\begin{center}
\textit{Figure \ref{fig:EigenvalueFig}: Sloppy model eigenvalues}
\end{center}

The PCA shown in Figure~\ref{fig:EigenvalueFig} column 
\emph{SP PCA} was produced after taking
logarithms of the parameter values that von Dassow \emph{et al.} used in
their analysis. This measures parameter fluctuations in terms of fractional
changes in parameter, rather than absolute sizes of fluctuations -- allowing
fluctuations in parameters with different units, for example, to be compared.
The parameters used in column~\emph{SP} were chosen (logarithmic or otherwise)
as defined by the original authors. Taking logarithms and/or changing units
does not typically change the qualitative spectra of sloppy models, as their
spectra already span so many decades.

\section*{Chemotype robustness and sloppiness}

\begin{center}
\textit{Derivation of robustness equation}
\end{center}
~\\
In the main text (MT), the robustness is defined as
\begin{equation}
R_c = \prod_{\lambda_n>\lambda_{crit}} \sqrt{\frac{\lambda_{crit}}{\lambda_n}}. \tag{MT \ref{eq:ChemotypeRobustness}}
\end{equation}
We now proceed to derive this result.  We measure robustness as the fraction of mutations of a given size $\delta$ in $C$ (chemotype space) that do not change the behavior beyond a given threshold (survival after a mutation), which we designate as an $\epsilon$-ball 
around the optimum in $D$ (dynatype space).  Therefore we want
an estimate of the fraction of the $\delta$-ball 
in $C$ that maps into the $\epsilon$-ball in $D$.  It is difficult to 
calculate this geometrically, since we would need to find the volume of an ellipsoid intersecting a sphere.  Fortunately, for sloppy systems, the $\lambda_i$ vary over many orders of magnitude, so we can simplify the calculation by smearing the $\delta$-ball and $\epsilon$-ball into Gaussians.  Namely, we say a mutation $\Delta\btheta$ in $C$ has probability 
$e^{-(\Delta\bthetaSmall)^2/2\delta^2}/(\sqrt{2\pi}\delta)^N$, 
and the probability of ``survival'' in $D$ is given by $e^{-r^2/2\epsilon^2}$.   We then measure the robustness as the overall probability $P ( \delta, \epsilon )$ of surviving after a mutation:
\begin{widetext} 	
\begin{align}
R_c &= P \left( \delta, \epsilon \right) \nonumber\\
    &= \left(\frac{1}{\sqrt{2 \pi}\delta}\right)^N \int_{C} \mathrm{d\Delta\btheta} \, \exp(-(\Delta\btheta)^2/2\delta^2) \exp(-(\Delta\btheta)^T J^T J (\Delta\btheta) / 2 \epsilon^2) \nonumber\\
   &= \prod_n \frac{1}{\sqrt{1 + \lambda_n \, \delta^2 / \epsilon^2}}. 
\end{align}
\end{widetext}		
For sloppy systems, $\lambda$ varies over many orders of magnitude.  Notice that if $\lambda_n \ll \epsilon^2 /\delta^2$, its component in the product will be 
close to 1, and if $\lambda_n \gg \epsilon^2 /\delta^2$, we can approximate the 
components in the product as $\sqrt{\epsilon^2/{\delta^2 \lambda_n}}$. 
Therefore, using our definition 
$\lambda_{crit} \equiv \epsilon^2/ \delta^2$ we can approximate this formula
as: 
\begin{equation}
R_c  \approx \prod_{\lambda_n > \epsilon^2/\delta^2} \sqrt{\frac{\epsilon^2}{\delta^2 \lambda_n}} 
  = \prod_{\lambda_n>\lambda_{crit}} \sqrt{\frac{\lambda_{crit}}{\lambda_n}},
\end{equation}
with small corrections for eigenvalues $\lambda_n\approx\epsilon^2/\delta^2$.
Since this result agrees with the ``slab'' argument given in the main text for hard walls, we see that hard $\epsilon$-balls and hard $\delta$-balls will have approximately the same amount of overlap as Gaussians.  

\section*{Robustness, evolvability, and sloppiness}
\begin{center}
\textit{Derivation of chemotype evolvability}
\end{center}

In the main text, we provide a formula for the ``maximum fitness change among mutations of size $\delta$ in chemotype space''
\begin{equation}
e_c(\bF, \btheta) =\sqrt{\bF^TJJ^T\bF}\delta \tag{MT \ref{eq:ChemotypeEvolvability}}
\end{equation}
which we derive here using a Lagrange multiplier.
To derive this, we use the definition of the chemotype evolvability as the
maximum response $\br \cdot \bF$ in $R$ for moves in $C$ of size 
$|\Delta\btheta|=\delta$:
\begin{equation}
e_c(\bF, \btheta) = \max_{|\Delta\bthetaSmall|=\delta}(\br \cdot \bF).
\end{equation} 
Next, notice that
\begin{equation}              
\br\cdot\bF = (J\Delta\btheta)\cdot\bF = \sum_i \sum_{\alpha} F_i J_{i \alpha} \Delta\theta_{\alpha}.
\end{equation}
We find the optimal $\Delta \btheta$ using a Lagrange multiplier $\Lambda$.  With $(\Delta\btheta)^2=\delta^2$ as our constraint, we maximize
\begin{equation}
F_i J_{i\alpha}\Delta\theta_{\alpha} + \Lambda ((\Delta\btheta)^2-\delta^2)
= F_i J_{i\alpha}\Delta\theta_{\alpha} + \Lambda (\Delta\theta_\beta \Delta\theta_\beta-\delta^2)
\end{equation}
where we use the Einstein summation convention (summing over repeated indices).
Differentiating with respect to $\Delta\theta_{\alpha}$, we can find the 
change $\Delta\theta^{max}$ giving the maximum response:
\begin{equation}
\Delta\theta^{max}_\alpha = \frac{F_j J_{j\alpha}}{2\Lambda}
\end{equation}
and hence
\begin{equation}
(\Delta\btheta^{max})^2 = 
	\frac{F_i J_{i\alpha} J_{j\alpha} F_j}{4\Lambda^2} = \delta^2,
\end{equation}
which implies
\begin{equation}
\Lambda^2 = \frac{\bF^T J J^T \bF}{4 \delta^2}.
\end{equation}
Therefore, the evolvability is:
\begin{align}
e_c(\bF, \btheta) &= F_i J_{i \alpha} \Delta\theta^{max}_{\alpha} 
			= \frac{F_i J_{i\alpha} F_j J_{j\alpha}}{2\Lambda} \\
		  &= \frac{\bF^T J J^T \bF}{\sqrt{\bF^T J J^T \bF}} \, \delta
                   = \sqrt{\bF^TJJ^T\bF} \, \delta. \nonumber
\end{align}

\begin{center}
\textit{RMS dynatype evolvability}
\end{center}

In Equation~(\ref{eq:Ec}), to measure overall evolvability, we defined $E_c(\btheta_\alpha)$ as a root-mean-square (RMS) average over a uniform (hyper)spherical distribution of environmental forces $\bF$ in dynatype space.  We use the RMS $\sqrt{\langle e_c(\bF,\btheta_\alpha)^2 \rangle}$ rather than the average $\langle e_c(\bF,\theta_\alpha) \rangle$ because the RMS definition has an elegant result in terms of the eigenvalues $\lambda_i$ of $J^TJ$:
\begin{align}
E_c(\btheta_\alpha)^2 &= \langle e_c(\bF,\btheta_\alpha)^2 \rangle_\bF = \langle \bF^T J J^T \bF \delta^2 \rangle_\bF \nonumber\\
                      &= \sum_i \frac{ \int \lambda_i F_i^2 \, \mathrm{d}^N\bF }{ \int \mathrm{d}^N\bF} \, \delta^2 \nonumber\\
		      &= \sum_i \lambda_i \langle F_i^2 \rangle \, \delta^2
		      = \frac{\sum_i \lambda_i \langle \bF^2 \rangle}{N} \, \delta^2 \nonumber\\
                      &= \frac{Tr(J^T J)\langle \bF^2 \rangle}{N} \, \delta^2 
                      \approx \frac{Tr(H)\langle \bF^2 \rangle}{N} \delta^2.
\end{align}
Therefore, the overall evolvability is directly related to the trace of the Hessian:
\begin{equation}
E_c(\btheta_\alpha) = \sqrt{\frac{Tr(H)\langle \bF^2 \rangle}{N}} \, \delta. \tag{MT 5}
\end{equation}

Our measures of robustness and evolvability depend upon our level of 
description, just as for Wagner's genotype and phenotype evolvabilities of
RNA sequences (Ref.~\cite{Wag08} of the main text).  Our choice of an
isotropic distribution of 
selective dynatype forces $\bF$ is not intended as an accurate representation
of actual selective forces at the phenotype level, but as an exhaustive
study of all possible forces at the dynatype level of description. 

Information 
about phenotypic selective pressures might suggest a different distribution
of dynatype forces $\bF$. Indeed, this formalism provides a mechanism 
for coupling maps across scales, which is an important unsolved problem.
Just as the genotype-to-chemotype ($G \to C$) and
chemotype-to-dynatype ($C \to D$) maps are many-to-one, so is the
dynatype-to-phenotype map ($D \to P$).  
In the segment polarity model,
for example, one might construe the phenotype as the steady-state
pattern, whereas the dynatype will include information about all
transient paths to that steady state. 
This is also closely analogous to measuring evolvability of 
RNA sequences by counting distinct folded structures (Ref.~\cite{Wag08} of the 
main text),
as many different structures may be equally nonfunctional at the higher
level of biological phenotype.  Ultimately, understanding the nature
of the complex $D \to P$ maps will be required to estimate
evolvability using more realistic distributions of selective dynatypic
forces $\bF$.

\begin{center}
\textit{Figure \ref{fig:Evolvability}: Evolvability and robustness in a 
sloppy system}
\end{center}

When calculating the chemotype robustness $R_c$,
we have a choice to make for the value of $\lambda_{crit}$ 
(see Equation~\ref{eq:ChemotypeRobustness}).  This choice corresponds to setting 
the ratio of the
size of acceptable changes in dynatype $\epsilon$ to the typical size of mutations $\delta$ in chemotype space: $\lambda_{crit}=\epsilon^2/ \delta^2$.  
Equivalently, $\lambda_{crit}$ sets a cutoff between stiff 
and sloppy eigenvalues, since we assume that, in $D$ space, 
the image of the $\delta$-ball fully overlaps with the $\epsilon$-ball in 
sloppy directions (with eigenvalues below $\lambda_{crit}$), and it extends
far beyond the edge of the $\epsilon$-ball in stiff directions 
(with eigenvalues above $\lambda_{crit}$).

In calculating $R_c$ for the inset of Figure~\ref{fig:Evolvability} in the
main text, 
we chose $\lambda_{crit}$ as the fourth stiffest eigenvalue of $J^T J$ at
the best fit parameters.  This matches with the idea that there are only a
few stiff directions that appreciably constrain parameters in chemotype
space: the eigenvalues are spaced by roughly factors of three 
(Figure~\ref{fig:EigenvalueFig}), meaning mutations in sloppier directions
in chemotype space quickly become irrelevant in dynatype space. 
The choice of $\lambda_{crit}$ within
a reasonable range (between, say, the second stiffest and eighth stiffest
eigenvalue of $J^T J$) does not
qualitatively change the plot of evolvability vs.~robustness.

\end{document}